\documentclass[a4paper,noarxiv,twocolumn]{quantumarticle}
\usepackage[utf8]{inputenc}
\usepackage[english]{babel}
\usepackage[T1]{fontenc}

\usepackage{lipsum}
\usepackage{mdframed}
\usepackage{booktabs}
\usepackage{longtable}

\usepackage{bm}
\usepackage{graphicx}
\usepackage{amsmath}
\usepackage{amsfonts}
\usepackage{amssymb}
\usepackage{amsthm}
\usepackage{amstext}
\usepackage{amsbsy}
\usepackage{amsopn}
\usepackage{amscd}
\usepackage{amsxtra}
\usepackage[colorlinks=true,allcolors=blue]{hyperref}
\usepackage{color}
\usepackage{soul}
\usepackage[dvipsnames]{xcolor}

\usepackage{ulem}

\def\phi{\varphi}
\def\epsilon{\varepsilon}

\def\Id{\textrm 1\!\!\!\!1}

\newcommand{\Mc}[1]{\mathcal{#1}}

\newcommand{\setR}{\mathbb{R}}

\newcommand{\setC}{\mathbb{C}}

\newcommand{\bra}[1]{\langle #1 |}
\newcommand{\ket}[1]{| #1 \rangle }
\newcommand{\braket}[2]{\langle #1| #2 \rangle }

\everymath{\displaystyle}

\begin{document}

\title{Quantum States Seen by a Probe: Partial Trace Over a Region of Space}





\author{Quentin Ansel}
\email{quentin.ansel@univ-fcomte.fr}
\affiliation{Institut UTINAM, CNRS UMR 6213, Universit\'{e} Bourgogne Franche-Comt\'{e}, Observatoire des Sciences de l'Univers THETA, 41 bis avenue de l'Observatoire, F-25010 Besan\c{c}on, France}
\orcid{0000-0002-4594-5978}

\date{\today}

\begin{abstract}
The partial trace operation is usually considered in composite quantum systems, to reduce the state on a single subsystem. This operation has a key role in the decoherence effect and quantum measurements. However, partial trace operations can be defined in more generic situations. In particular, it can be used to restrict a quantum state (for a single or several quantum entities) on a specific region of space, the rest of the universe being treated as an environment. The reduced state is then interpreted as the state that can be detected by an ideal probe with a limited spatial extent. In this paper, such an operation is investigated for systems defined on a Fock Hilbert space. A generic expression of the reduced density matrix is computed, and it is applied to several case studies: eigenstates of the number operator, coherent states, and thermal states. These states admit very different behaviors. In particular, (i) a decoherence effect happens on eigenstates of the number operator (ii) coherent or thermal states remain coherent or thermal, but with an amplitude/temperature reduced non-trivially by the overlap between the field and the region of interest.
\end{abstract}

\maketitle
\tableofcontents
\section{Introduction}

The partial trace is a key concept in the theory of open quantum system~\cite{breuer_theory_2007,jacobs_quantum_2014,gardiner_quantum_2004}. It aims at describing the state of a quantum system in interaction with an environment. This latter does not have specific degrees of freedom in the system state, it is "traced out". This operation is defined when the Hilbert space $\Mc H$ of the full system can be expressed as a tensor product of two Hilbert spaces $\Mc H_s$ and $\Mc H_e$, associated respectively with the system and the environment. Then the reduced state (density matrix) is defined by
\begin{equation}
\hat \rho_s = \text{Tr}_e [\hat \rho] = \sum_{ijk} \bra{i,k}\hat \rho \ket{j,k} \ket{i}\bra{j}
\end{equation}
where $\ket{i,k} \in \Mc H = \Mc H_s \otimes \Mc H_e$ are level states $i$ and $k$ of the system and the environment, and $\ket{i} \in \Mc H_s$ are states of the system only.

A typical application of such an operation is the computation of an atomic state in interaction with the electromagnetic field~\cite{breuer_theory_2007}. By using a time-dependent state for these two entities and the partial trace, the time evolution of the atomic state alone can be computed. The electromagnetic field is then present through dynamical properties. The main advantage of such an operation is that the dimension of the system can be considerably reduced, and it is usually the starting point for both analytic and numerical computations.

In this example, the system and the environment are straightforwardly identified since they are two distinct physical entities. However, a similar idea can be applied to more subtle situations. Among them, there is the case of a complex quantum system, with a very large Hilbert space, for which few energy levels are isolated, and the other levels are treated as an environment (levels of the environment interact rather weakly with the levels of interest). In such a case, a partial trace is also relevant, albeit more subtle to establish (using e.g., effective systems~\cite{azouit_adiabatic_2016}). Until now, the states have referred to function spread over a potentially wide region, and the partial trace does not necessarily change the size of the region. \textit{One can also wonder what happens to the state of a system when it is restricted to a small region of space, potentially smaller than the one in which it is located}. Again, a partial trace is the natural mathematical operation to use in such a circumstance. This question is usually not considered in standard quantum mechanical problems because we can avoid this issue. However, it tends to become necessary in quantum field in curved space-time~\cite{10.1063/1.4939955} or quantum gravity (mostly in loop quantum gravity or in the spinfoam formalism~\cite{rovelli_covariant_2014,ansel_model_2021}) since the geometry of space is not flat, and it is useful to restrict the study to a small region that can be modeled accurately with a minimum of degrees of freedom. 

It can also be relevant for the study of a quantum state in two non-inertial reference frames. In such a case, there are causally disconnected regions, and after the change of frame, these regions are traced out, to keep only the one causally connected with the observer. This idea is at the origin of the Hawking~\cite{hawking_particle_1975} and the Unruh~\cite{RevModPhys.80.787} effects. With causally disconnected regions, the partial trace is conceptually simple to establish, but the very same operation can be carried out for any region of space, even with causally connected ones. 
The construction of this kind of state is relevant in the context of quantum measurements. A probe can interact with a wave packet whose spatial extent is larger than the probe size. In this case, the detector cannot resolve the entire field state. A state tomography~\cite{lundeen2009tomography,cooper2014local} can only lead to a state restricted to a given region of space~\cite{PhysRevLett.87.050402}. Then, the reduced state is interpreted as the state "seen" by a detector with a finite (and small) spatial extent. 

These notes aim to investigate the partial trace of a quantum state over a (small) region of space. Here, space-time is treated classically. The idea is to look at the effect of this partial trace in some very simple examples, to provide well-established properties and concepts that could be reused in more complicated situations. Even though the examples are very standard, the results are far from being obvious. They can be directly reused in practical applications, like photon counting~\cite{srinivas1981photon,fox2006quantum,haroche_exploring_2006,polino_photonic_2020}, where the detector is not well adapted to the size of the incoming photons.

The paper is organized as follows. First, a few elementary materials on Fock spaces and creation/annihilation operators are given. Then, a general formula for the partial trace is derived. Finally, the formula is applied to 3 types of quantum states, $n$-excitation states, coherent states, and thermal states.

\section{Preliminary Materials}

To define a partial trace operation over a region of space, several key concepts on Fock spaces must be recalled~\cite{Zeidler2009,honegger2015photons}. To keep the discussion as simple as possible, the most standard tools of quantum fields are used. Specific features of algebraic quantum fields theory are not necessary, and thus, it is not used in the following\footnote{Just to fix ideas, from the point of view of algebraic quantum field theory, the problem of constructing a reduced state can be written as follows: one would like to determine a state $\hat \rho_0 \in \mathfrak{A}(O_0 \subseteq O )$ from the state $\hat \rho \in \mathfrak{A}(O)$, such that $\forall \hat A \in \mathfrak{A}(O_0) $, $\text{Tr}[\hat \rho_0 \hat A] = \text{Tr}[\hat \rho \hat A]$.}.

One of the first key points is that contrary to a quite widely believed idea, plane waves are not necessary to define the Fock space of a quantum field. This latter can be defined using any arbitrary function of $\Mc V =L_2(\setC^N,d^3x)$, $\Mc V$ being the Hilbert space used to define the Fock space $\Mc F = \bigoplus_{n=0}^\infty S \Mc V^{\otimes n}$, with $S$ an operator which symmetrizes or antisymmetrizes a tensor (in the following, a focus is made on symmetrized Fock spaces). A direct consequence is that creation and annihilation operators can be defined for arbitrary elements of $\Mc V$, which means that for a quantum system, such as the electromagnetic field, these operators can be constructed for any classical field shape. To fix the notation, we define $\hat a_q ^\dagger$ and $\hat a_q$ the operators that create or destroy a quantum excitation for a field $q \in \Mc V$. Since $q$ belongs to a Hilbert space, it can be decomposed into a sum of other fields, i.e. $q=\sum_i q_i \eta_i$, $q_i \in \setC$ and $\eta_i \in \Mc V$. Then, it is possible to define creation/annihilation operators for each $\eta_i$, noted $\hat a^\dagger _i$ and $\hat a_i$. The operators must follow these properties:
\begin{align}
\hat a_q^\dagger& = \sum_i q_i \hat a_i^\dagger, \\
\hat a_q& = \sum_i q_i^* \hat a_i, \\
[\hat a_i ,\hat a_j^\dagger] &= \braket{\eta_i}{\eta_j} \Id_{\Mc F},\\
[\hat a_i, \hat a_j]&=[\hat a_i^\dagger, \hat a_j^\dagger]=0,
\end{align}
where $\braket{\eta_i}{\eta_j}$ is the scalar product in $\Mc V$ and $ \Id_{\Mc F}$ is the identity operator in $\Mc F$. In fact, these properties are specific cases of Bogoliobov transformations~\cite{ruijsenaars_bogoliubov_1978,bogoliubov_1997}. They are extensively used in condensed matter physics and quantum field theory in curved space-time. In particular, the Bogoliobov transformation is at the core of the Hawking radiation or the Unruh effetcs~\cite{hawking_particle_1975,RevModPhys.80.787}.

\begin{figure*}
\includegraphics[width=\textwidth]{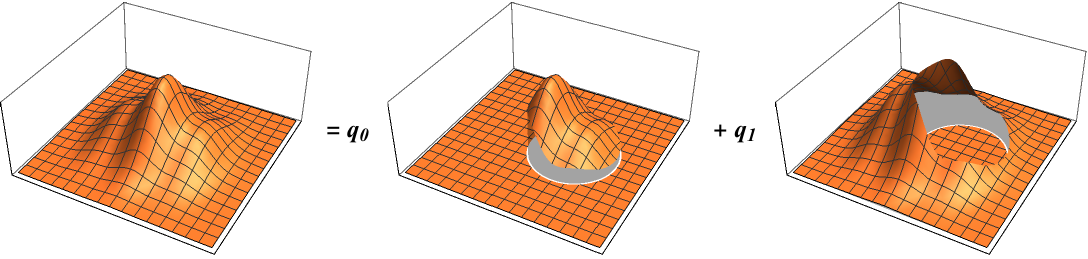}
\caption{Illustration of the splitting of the mode $q$ into two functions whose supports are disjoint.}
\label{fig:mode_splitting}
\end{figure*}

With these properties at hand, it is straightforward to decompose a field operator $\hat a_q ^\dagger$ into two field operators, each one associated with a different region of space. We define~\cite{PhysRevA.36.1955}
\begin{equation}
q = q_0 \eta_0 + q_1 \eta_1
\end{equation}
The function $\eta_0$ has a compact support on $\Mc R \subset \setR^3$, and the function $\eta_1$ has a support on $\bar{\Mc R}$. The situation is illustrated in Fig.~\ref{fig:mode_splitting}. Moreover, one must have 
\begin{equation}
q^2_0 + q_1 ^2 =1.
\end{equation}
In the following, we restrict $q_0$ and $q_1$ to real numbers, but the case with complex numbers can be easily deduced.

\section{The Partial Trace Over a Region of Space}

Using the materials presented in the previous section, it is now possible to establish a rather general formula for the partial trace of a quantum state over a region of space.

First, let us consider a pure state of the form
\begin{equation}
\ket{\psi_q} = \sum_n \frac{\psi_n}{\sqrt{n!}}(\hat a^\dagger_q)^m \ket{0}.
\end{equation}
The density matrix is then:
\begin{equation}
\hat \rho = \ket{\psi_q}\bra{\psi_q} = \sum_{nm} \frac{\psi_n \psi_m^*}{\sqrt{n! m!}} (\hat a^\dagger_q)^n \ket{0} \bra{0} (\hat a_q)^n 
\end{equation}
Next, $\hat a_q^\dagger$ and $\hat a_q$ can be replaced by $q_0 \hat a_0^\dagger + q_1 \hat a_1 ^\dagger$ and $q_0 \hat a_0 + q_1 \hat a_1 $, respectively. Using also the binomial theorem, one arrives at:

\begin{equation}
\begin{split}
\hat \rho =& \sum_{nm} \psi_n \psi_m^* \left(\sum_{k=0}^n \frac{\sqrt{n!}}{k! (n-k)!)} (q_1 \hat a_1^\dagger)^k (q_0 \hat a^\dagger_0)^{n-k} \right) \ket{0} \\
& \times \bra{0} \left( \sum_{l=0}^m \frac{\sqrt{m!}}{l! (m-l)!)} (q_1 \hat a_1)^l (q_0 \hat a_0)^{m-l}\right).
\end{split}
\end{equation}
Using this last expression, the reduced density matrix $\hat \rho_0$ where the region "1" is traced out can be computed. It is given by\footnote{A similar formula is found in \cite{PhysRevLett.87.050402}, but it remains vague, and no explicit calculations were performed.}
\begin{equation}
\hat \rho_0 = \text{Tr}_1 [\hat \rho] = \sum_{ijk} \bra{i,k}\hat \rho \ket{j,k} \ket{i}\bra{j}
\end{equation}
where, the state $\ket{i,k}$ denotes a state with $i$ excitations in the region 0, and $k$ excitations in the region 1. Also, the state $\ket{i}$ corresponds to a state with $i$ excitations in the region 0, the region 1 being traced out. Using the orthogonality relation, it is straightforward to determine the following expression for the matrix elements of $\hat \rho_0$:
\begin{equation}
\bra{i} \hat \rho_0 \ket{j} = \sum_{o}^\infty
 \psi_{o+i} \psi_{o+j}^* \frac{ \sqrt{(o+i)!(o+j)!}}{o! \sqrt{i!j!}}q_1^{2o} q_0^{j+i},
 \label{eq:partial_trace_general}
\end{equation}
for $j \geq i$. In the case $j<i$, the result is simply the complex conjugated of Eq.~\eqref{eq:partial_trace_general}.

This result can be directly used for the case where $\hat \rho$ is a mixed state
\begin{equation}
\hat \rho = \sum_n p_n \ket{\psi_{q,n}} \bra{\psi_{q,n}}.
\end{equation}
By linearity of the partial trace, one gets
\begin{equation}
\hat \rho_0 = \sum_n p_n \text{Tr}_1[ \ket{\psi_{q,n}}\bra{\psi_{q,n}}], 
\end{equation}
with $\text{Tr}_1[ \ket{\psi_{q,n}}\bra{\psi_{q,n}}]$ evaluated using Eq.~\eqref{eq:partial_trace_general}.

\section{Study of a Few Examples}

From the general expression Eq.~\eqref{eq:partial_trace_general}, it is possible to compute explicitly the reduced density matrix with a specific choice of state $\ket{\psi_q}$. In the following paragraphs, `$n$-excitations, coherent and thermal states are investigated.

\subsection{$n$-excitation states}

A $n$-excitation state is defined by $\ket{\psi_ q} = \ket{n_q}$. It is an eigenstate of the number operator, i.e. $\hat a_q^\dagger \hat a_q \ket{n_q} = n \ket{n_q}$. With such a state, the density matrix is $\hat \rho = \ket{n}\bra{n}$. Now, using Eq.~\eqref{eq:partial_trace_general} to compute the matrix elements of the reduced state, one sees that $\psi_{o+i} = \delta_{o+i,n}$ and $\psi^*_{o+j}= \delta_{o+j,n}$. These Kronecker symbols imply $i=j$, and only states smaller than $n$ are populated. The final result is
\begin{equation}
\begin{split}
\hat \rho_0 &= \sum_{i=0}^n \frac{n!}{(n-i)! i!}q_1^{2(n-i)}q_0^{2i} \ket{i} \bra{i} \\
 &= \sum_{i=0}^n \frac{n!}{(n-i)! i!}(1-q_0^2)^{2(n-i)}(q_0^{2})^i \ket{i} \bra{i} \\
& = \sum_{i=0}^n B(n,q_0^2) \ket{i} \bra{i}
\end{split}
\label{eq:reduced_state_n-photons}
\end{equation}
This state is generally not pure. It corresponds to a statistical mixture where the probability to measure $i \leq n$ excitation follows a binomial distribution $B(n,q_0^2)$. The Eq.~\eqref{eq:reduced_state_n-photons} agrees with a standard result in the field of photon counting~\cite{gardiner_quantum_2004,srinivas1981photon,haroche_exploring_2006,fox2006quantum}. It is usually derived by computing the probability to detect eigenvalues of the number operator, without taking into account the system coherence. Here, the result is a natural consequence of the partial trace operation.

\begin{figure}
\includegraphics[width=\columnwidth]{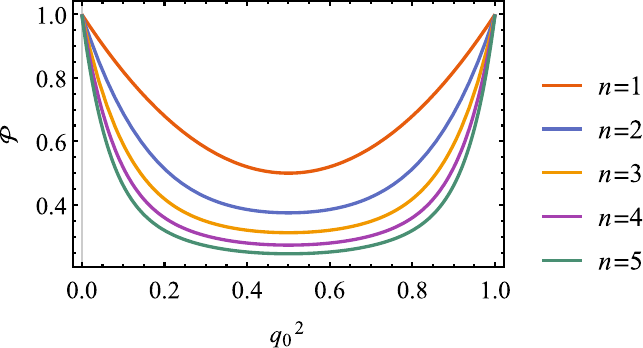}
\caption{Purity $\Mc P$ as a function of $q_0 ^2$ for reduced $n$-excitations states, for $n$ in the range $[1,5]$.}
\label{fig:purity_n_photons}
\end{figure}

The purity of the quantum state can be studied by computing
\begin{equation}
\Mc P=\text{Tr}[\hat \rho_0^2] = \sum_{i=0}^n B(n,q_0^2)^2.
\end{equation}
It is plotted as a function of $q_0^2$ in Fig.~\ref{fig:purity_n_photons}, for some values of $n$. 
$\Mc P=1$ only when $q_0 = 0$ or $1$. These are the only cases where the state is pure, which corresponds to situations where excitations are completely outside or contained into the region $\Mc R$. Otherwise,$\Mc P<1$. The minimum value is reached when $q_0^2 = q_1^2 = 1/2$.

This section is finished with a short study of the effect of the partial trace on an entangled state. For simplicity, the initial state is chosen to be a "Schröndinger cat" state $\ket{\psi_q} = \tfrac{1}{\sqrt{2}}(\ket{0_q} + \ket{1_q}$. 

The reduced density matrix is then
\begin{equation}
\hat \rho_0 = \left(1- \frac{q_0^2}{2}\right) \ket{0} \bra{0} + \frac{q_0^2}{2} \ket{1} \bra{1} + \frac{q_0^2}{2} \left( \ket{0} \bra{1} + \ket{1} \bra{0} \right).
\end{equation}
The purity of this state is given by $\Mc P = \tfrac{1}{2}(2-q_0^2 + q_0^4)$. It is equal to the purity of a single excitation state, plotted in Fig.~\ref{fig:purity_n_photons}. To conclude, the partial trace operation tends to reduce the entanglement of the system. This is quite a common behavior encountered in open quantum system theory~\cite{breuer_theory_2007}.

\subsection{Coherent state}

A coherent state parameterized by the complex number $\alpha$ is defined by~\cite{gardiner_quantum_2004}:
\begin{equation}
\ket{\psi_q} = e^{- |\alpha|^2/2} \sum_{n=0}^\infty \frac{\alpha ^n}{\sqrt{n!}} (\hat a_q^\dagger)^n \ket{0}.
\end{equation}
When this state is inserted in Eq.~\eqref{eq:partial_trace_general}, one gets
\begin{equation}
\begin{split}
\bra{i} \hat \rho_0 \ket{j}& =e^{- |\alpha|^2} \sum_{o}^\infty
 \alpha^{o+i} {\alpha^*}^{o+j} \frac{ 1}{o! \sqrt{i!j!}}q_1^{2o} q_0^{j+i} \\
 & = e^{- |\alpha|^2} \frac{(q_0\alpha)^{i} (q_0\alpha^*)^{j}}{ \sqrt{i!j!}} \sum_{o=0}^\infty \frac{|\alpha|^{2o} q_1 ^{2o}}{o!} \\
 &= e^{- |\alpha|^2} \frac{(q_0\alpha)^{i} (q_0\alpha^*)^{j}}{ \sqrt{i!j!}} e^{|\alpha|^2 q_1^2} \\
 &= e^{- q_0^2|\alpha|^2} \frac{(q_0\alpha)^{i} (q_0\alpha^*)^{j}}{ \sqrt{i!j!}}.
 \end{split}
\end{equation}
The last line is determined using the condition $q_1^2 = 1 - q_0^2$. A key observation is that these matrix elements correspond to the ones of a pure coherent state, parameterized by the complex number $q_0 \alpha$. Then, the reduced state can be rewritten into
\begin{equation}
\hat \rho_0 = \ket{q_0 \alpha} \bra{q_0 \alpha}.
\end{equation}
To conclude, a reduced coherent state remains a coherent state, but with an amplitude rescaled by the fraction of the field contained into the region $\Mc R$. In contrast with $n$-excitations states, the reduced state is always pure, and no decoherence effect can be observed.

\subsection{Thermal state}

\begin{figure}[t!]


\includegraphics[width=\columnwidth]{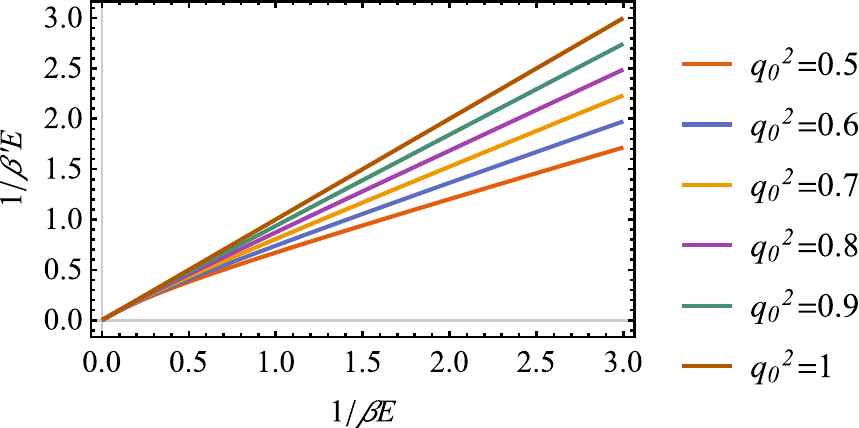}
\caption{
Normalized temperature $1/\beta'E$ of the reduced state as a function of the normalized temperature $1/\beta E$ of the thermal state, for different values of $q_0^2$.}
\label{fig:thermal_state}
\end{figure}
In this section, the state is assumed thermal with respect to the number operator $\hat N_q = \hat a^\dagger_q \hat a_q$. It is given by the density matrix~\cite{breuer_theory_2007}
\begin{equation}
\hat \rho = \Mc N_\beta \sum_{n=0}^\infty e^{-\beta E n} \ket{n} \bra{n},
\label{eq:def_thermal_state}
\end{equation}
where $\beta$ is the inverse temperature, $E$ the difference of energy between two levels, and $\Mc N_\beta = 1- e^{\beta E}$ is a normalization coefficient. The reduced state can be determined using Eq.~\eqref{eq:reduced_state_n-photons}. The result is

\begin{equation}
\begin{split}
\hat \rho_0 & = \Mc N_\beta \sum_{n=0}^\infty e^{-\beta E n} \sum_{i=0}^n \frac{n!}{(n-i)! i!}q_1^{2(n-i)}q_0^{2i} \ket{i} \bra{i}\\
&= \Mc N_\beta \sum_{k,m =0}^\infty e^{-\beta E (k+m)} \frac{(k+m)!}{k! m!} (1-q_0^2)^{k} q_0^{2m} \ket{m}\bra{m}\\
& = \sum_{m=0}^\infty \frac{(e^{\beta E} - 1) e^{-\beta E (m + 1)} q_0^{2m}}{(e^{-\beta E} (q_0^2-1)+1)^{m+1}}\ket{m}\bra{m}\\
&= \Mc N_{\beta'} \sum_{m=0}^\infty e^{-\beta' E m} \ket{m}\bra{m}
\end{split}
\label{eq:reduced_thermal_state}
\end{equation}
where between the first and the second line, the density matrix is projected on the states $\ket{m}$, and the sum over $n$ is redefined into a sum over k. Moreover, in the last line, one has defined
\begin{equation}
\beta'= \frac{1}{E} \ln \left( \frac{q_0^2 + e^{\beta E} - 1}{q_0^2}\right).
\end{equation}
The reduced state is therefore a thermal state for which the temperature is modified non-trivially. The relation between the initial temperature and the temperature of the reduced state is plotted in Fig.~\ref{fig:thermal_state}. The key observation is that the smaller $q_0^2$, the smaller the temperature of the reduced state. This effect is mostly noticeable for large initial temperatures. For small initial temperatures, the effect is very weak. Moreover, we can show that $\beta ' \rightarrow 0$ when $q_0^2 \rightarrow 0$, and $\beta ' \rightarrow \beta $ when $q_0^2 \rightarrow 1$ (which is expected since in these limits, $q$ is entirely outside or contained in the region $\Mc R$).

\section{Conclusion}

In these notes, a general formula for the computation of the partial trace over a region of space is derived for a quantum system defined on a Fock space. The calculation assumes the existence of two regions, one is the region of interest, and the other is treated as an environment. The formula is then applied to 3 types of quantum states. Very different properties of the reduced states are observed. In the case of $n$-excitation (pure) states, the reduced state is a statistical mixture of states with $i\leq n$ excitations. We thus observe a very typical effect of open quantum systems, which can be interpreted as a decoherence effect. For coherent and thermal states, the reduced state behaves differently: it preserves the nature of the state (coherent or thermal), but the amplitude or the temperature is modified. These modified parameters depend non-trivially on $q_0^2$ (the overlap between the field and the region of interest).

Coherent and thermal states have well-known properties that allow us to study the transition from the quantum to the classical regime. These results provide yet another property: \textit{the scale invariance}. They can be restricted to any small region of space (in the limit of validity of a continuous space-time), and their properties are preserved. The opposite vision is also true, if one observes a coherent or a thermal state in a small region of space, these states are potentially included in a widespread similar state.

These properties can be used directly in quantum metrology~\cite{polino_photonic_2020}, but they are also relevant in quantum gravity for which it can be arduous to define semi-classical states. Several generalizations of coherent states exist (like for coherent spin-networks~\cite{rovelli_covariant_2014}), but it is not obvious which properties of standard Perelomov's coherent states must be conserved or which ones can be relaxed. Here, the scale invariance could be a very useful property since it may allow us to construct very interesting reduced states (e.g., stable under "Pachner move" in the spinfoam formalism~\cite{rovelli_covariant_2014,borissova2023lorentzian}).


\end{document}